# Photonic thermal diode enabled by surface polariton coupling in nanostructures

Lei Tang and Mathieu Francoeur[*]

*Radiative Energy Transfer Lab, Department of Mechanical Engineering, University of Utah, Salt Lake City, UT 84112, USA*

**Abstract.** A novel photonic thermal diode concept operating in the near field and capitalizing on the temperature-dependence of coupled surface polariton modes in nanostructures is proposed. The diode concept utilizes terminals made of the same material supporting surface polariton modes in the infrared, but with dissimilar structures. The specific diode design analyzed in this Letter involves a thin film and a bulk, both made of $3C$ silicon carbide, separated by a subwavelength vacuum gap. High rectification efficiency is obtained by tuning the antisymmetric resonant modes of the thin film, resulting from surface phonon-polariton coupling, on- and off-resonance with the resonant mode of the bulk as a function of the temperature bias direction. Rectification efficiency is investigated by varying structural parameters, namely the vacuum gap size, the dielectric function of the substrate onto which the film is coated, and the film thickness to gap size ratio. Calculations based on fluctuational electrodynamics reveal that high rectification efficiencies in the 80% to 87% range can be maintained in a wide temperature band ($\sim$ 700 K to 1000 K). The rectification efficiency of the proposed diode concept can be potentially further enhanced by investigating more complex nanostructures such as gratings and multilayered media.

---

[*] Corresponding author. Tel.: +1 801 581 5721, Fax: +1 801 585 9825
Email address: mfrancoeur@mech.utah.edu

A photonic thermal diode is a two-terminal device in which the magnitude of radiative heat flow depends on the direction of the temperature bias.[1-3] Photonic thermal diodes are contactless and their ability to transport heat preferentially along one direction could impact thermal management of electronic and optoelectronic devices, as well as the development of thermal transistors,[4] switches[5,6] and memories.[7] The performance of a thermal diode is typically quantified by the rectification efficiency, defined as $\eta(T_h, T_l) = \left| Q_f - Q_r \right| / \max(Q_f, Q_r)$, where $Q_f$ and $Q_r$ are heat rates for the same temperature bias $\Delta T = T_h - T_l$, but in different, forward and reverse, directions.[1,3] Conventional photonic thermal diodes, however, are limited to relatively small heat flow due to Planck's blackbody limit.[1,2,8-10] To circumvent this problem, Otey et al.[11] proposed a photonic thermal diode capitalizing on the near-field effects of thermal radiation, thus enabling heat flow to exceed the blackbody limit. In this paradigm, the two terminals, separated by a subwavelength vacuum gap, are made of different bulk materials supporting resonant surface polariton modes. Thermal rectification is induced by the dissimilar temperature-dependence of the terminal resonances. The thermal diode proposed in Ref. 11, made of $6H$ and $3C$ silicon carbide (SiC), resulted in a rectification efficiency of 29.1% (rectification factor, $R(T_h, T_l) = \left| Q_f - Q_r \right| / \min(Q_f, Q_r)$, of 0.41) for a 100-nm-thick vacuum gap. Similar diodes involving terminals made of dissimilar bulks,[12,13] thin films,[14,15] and metamaterials[16] have been designed. A rectification efficiency as high as 91% ($R$ value of 9.9) was reported in Ref. 12 between intrinsic silicon and silicon dioxide for a small vacuum gap thickness of 5 nm. Near-field based photonic thermal diodes capitalizing on the insulator-metal transition of vanadium dioxide ($VO_2$) have also been proposed.[17-20] A maximum rectification efficiency of 94% ($R$ value of 16) was reported with a one-dimensional grating of $VO_2$ for a 100-nm-thick vacuum gap.[19]

While this approach can lead to high rectification efficiency, such thermal diodes are limited to operating near the critical temperature of phase change materials.

All aforementioned near-field photonic thermal diode designs capitalize on asymmetric heat flow induced by terminals made of dissimilar materials. Yet, strong thermal rectification in the near field can also be obtained with terminals made of the same material, a topic that has been essentially unexplored. Only Zhu et al.[21] analyzed thermal rectification in a diode made of two subwavelength nanospheres made of $3C$-SiC of different sizes. Since both nanospheres are much smaller than the wavelength, their resonant frequencies are size independent and thus always on-resonance regardless of the temperature bias direction. In the two-nanosphere thermal diode, rectification is mediated by the different strengths of the resonance coupling constant as a function of the temperature bias direction. Such a diode can lead to a rectification efficiency as high as 84% ($R$ value of 5.14). Yet, the small amount of heat that can be transferred between nanospheres, proportional to their surface area, is a drawback of this concept.

In this Letter, an alternative and simple diode concept, leading to high rectification efficiency, is proposed where both terminals are made of the same material supporting surface polariton modes in the infrared. As originally proposed in Ref. 11, rectification is achieved by switching the terminals on- and off-resonance via the direction of the temperature bias. However, conversely to previous work, on- and off-resonance is obtained by capitalizing on the temperature-dependence of coupled surface polariton modes in nanostructures.

The proposed rectification concept is illustrated via the photonic thermal diode shown in Fig. 1 consisting of two planar structures, called terminals $A$ and $B$, separated by a subwavelength vacuum gap of thickness $d$. Both terminals are made of $3C$-SiC, supporting surface phonon-polaritons (SPhPs) in the infrared, but are structurally dissimilar in order to induce on- and off-resonance as a function of the temperature bias direction. Specifically, terminal $A$ is made of a SiC thin film coated on a low-emitting substrate ($\varepsilon_{A2} = \varepsilon'_{A2} + i\delta$, $\delta \to 0$). Terminal $A$ thus support two resonant modes, namely a high-frequency antisymmetric and a low-frequency symmetric mode, due to SPhP coupling within the thin film.[22] Terminal $B$ consists of a bulk (i.e., optically thick) layer of SiC, and therefore supports a single resonant mode. Rectification is achieved by tuning the antisymmetric mode of terminal $A$ on- and off-resonance with the mode of terminal $B$ as a function of the temperature bias direction. Note that SPhP resonance splitting is visible on the near-field thermal spectrum emitted by a film only when its thickness $t$ is smaller than the vacuum gap size $d$,[23,24] such that $t < d$ is a necessary condition to achieve thermal rectification. An advantage of using a nanostructure such as a thin film comes from the fact that nonresonant thermal emission of propagating and frustrated modes, detrimental to the rectification efficiency, is greatly reduced since it is proportional to the volume of the heat source.

[Insert Figure 1]

The total near-field radiative heat flux between terminals $A$ and $B$, derived from fluctuational electrodynamics,[25] is calculated as follows:[24]

$$q_{f,r} = \frac{1}{4\pi^2} \int_0^\infty \Delta\Theta(\omega, T_h, T_l)\, d\omega$$

$$\times \left[ \int_0^{k_0} k_\rho\, dk_\rho \sum_{\gamma=TE,TM} \frac{\left(1 - \left|R_A^\gamma\right|^2\right)\left(1 - \left|R_B^\gamma\right|^2\right)}{\left|1 - R_A^\gamma R_B^\gamma e^{2ik'_{z0}d}\right|^2} + 4 \int_{k_0}^\infty k_\rho\, dk_\rho\, e^{-2k''_{z0}d} \sum_{\gamma=TE,TM} \frac{\operatorname{Im}\left(R_A^\gamma\right)\operatorname{Im}\left(R_B^\gamma\right)}{\left|1 - R_A^\gamma R_B^\gamma e^{-2k''_{z0}d}\right|^2} \right] \quad (1)$$

where $k_{\rho}$ and $k_{zi}$ ($=k'_{zi}+ik''_{zi}$) are the wavevector components parallel and perpendicular to the layer interfaces, while $k_0$ is the magnitude of the wavevector in vacuum (= $\omega/c_0$). The mean energy of a Planck oscillator, $\Delta\Theta(\omega,T_h,T_l)$, is calculated as $\Theta(\omega,T_h)-\Theta(\omega,T_l)$. When the temperature bias is in the forward direction, terminals $A$ and $B$ are respectively at temperatures $T_h$ and $T_l$, and vice-versa for reverse temperature bias. In Eq. (1), the term $R_j^{\gamma}$ is the reflection coefficient of terminal $j$ in polarization state $\gamma$ (TE: transverse electric, TM: transverse magnetic). The reflection coefficient of terminal $A$ is calculated as $R_A^{\gamma}=\left(r_{0,A1}^{\gamma}+r_{A1,A2}^{\gamma}e^{2ik_{z1}t}\right)\Big/\left(1+r_{0,A1}^{\gamma}r_{A1,A2}^{\gamma}e^{2ik_{z1}t}\right)$, while the reflection coefficient of terminal $B$ is $R_B^{\gamma}=r_{0,B1}^{\gamma}$, where $r_{i,j}^{\gamma}$ denotes the Fresnel reflection coefficient at the interface $i,j$ in polarization state $\gamma$.[26] The temperature-dependent dielectric function model of 3$C$-SiC, taken from Ref. 21, is provided in the supplementary material.[27] Since both terminals have the same cross-sectional area, the rectification efficiency can be computed using heat fluxes, given by Eq. (1), rather than heat rates. In all results presented hereafter, the low temperature $T_l$ is always kept at 300 K, while the high temperature $T_h$ varies from 400 K to 1000 K.

Rectification efficiency is plotted against the high temperature $T_h$ and the gap size $d$ in Fig. 2(a) when the dielectric function of the substrate, $\varepsilon'_{A2}$, is equal to 1 and the film thickness to gap size ratio, $D = t/d$, is fixed at 0.1. The rectification efficiency drops drastically as the gap size increases. For instance, when $T_h = 900$ K, the average rectification efficiency is around 60% for $d$ values between 10 nm and 50 nm, and it is close to zero when the gap size reaches 500 nm. Such a behavior is to be expected, since SPhPs dominate radiative heat transfer in the extreme near field, up to a gap size of approximately 100 nm.[28,29] Here, it is interesting to note that a relatively high rectification efficiency ($\sim 60\%$) is obtained in a wide band of gap size ($\sim 10$ nm $- 50$ nm) and temperature ($\sim 700$ K $- 1000$ K).

SPhP dispersion relation can be used to explain the variation of rectification efficiency as a function of the temperature $T_h$. Since SiC is nonmagnetic, SPhPs can only be excited in TM polarization.[22] To account for cross-coupling of SPhPs between terminals $A$ and $B$, the dispersion relation is obtained by solving $1 - R_A^{TM} R_B^{TM} e^{-2k_{z0}''d} = 0$ in the electrostatic limit where $k_\rho >> k_0$, such that $k_{zj} \approx ik_\rho$.[24] SPhP dispersion relation as a function of the temperature $T_h$ is shown in Fig. 2(b) for the same parameters as in Fig. 2(a), except that the gap size is fixed at 10 nm. SPhP dispersion relation is calculated at the dominant wavevector $k_\rho \approx 1/d$ and by neglecting losses in the dielectric function of SiC.[24] As expected, the dispersion relation shows that three resonant modes occur in the diode. The frequencies $\omega_1$ and $\omega_3$ are respectively the antisymmetric and symmetric modes of terminal $A$ that are slightly modified, compared to the case of an isolated film, due to cross-coupling with terminal $B$. The frequency $\omega_2$ is the resonance of the SiC bulk that couples with terminal $A$. The modes $\omega_1$ and $\omega_2$ exhibit totally different behaviors in the forward- and reverse-biased scenarios. As the temperature increases, $\omega_1$ and $\omega_2$ approach each other in the forward case, while the inverse behavior is observed for the reverse bias. As such, a high rectification efficiency is obtained due to strong on- and off-resonance switching starting at a temperature of approximately 700 K.

[Insert Figure 2 here]

The impact of the substrate in terminal $A$ on the thermal rectification efficiency is analyzed in Fig. 3, where $D$ and $d$ are fixed at 0.1 and 10 nm, respectively. In the temperature band from 700

K to 1000 K, a high thermal rectification efficiency (> 80%) is maintained for various substrate dielectric functions from $\varepsilon'_{A2} = 2$ to 12. Here, the case $\varepsilon'_{A2} = 12$ approximates quite well the dielectric function of intrinsic silicon in the infrared. Compared to Fig. 2, the rectification efficiency is further enhanced by increasing the dielectric function of the substrate. As in the previous case, rectification still comes from switching $\omega_1$ and $\omega_2$ on- and off-resonance as a function of the temperature bias direction. SPhP dispersion relation displayed in Fig. 3(b) for $T_h$ = 900 K shows that both $\omega_1$ and $\omega_2$ are essentially insensitive to $\varepsilon'_{A2}$. The rectification efficiency enhancement as a function of $\varepsilon'_{A2}$ is due to the fact that the frequency of the mode $\omega_3$ slowly decreases as the dielectric function of the substrate increases. This can be better understood by inspecting the spectral distributions of radiative heat flux shown in Fig. S1 of supplemental material.[27] The low-frequency mode $\omega_3$ exhibits a low degree of spectral coherence, compared to the $\omega_1$ and $\omega_2$ modes, due to higher losses.[24] This leads to potential interaction between the $\omega_2$ and $\omega_3$ modes, negatively impacting the diode rectification efficiency (see Fig. S1(a)). By increasing the dielectric function of the substrate, the interaction between $\omega_2$ and $\omega_3$ decreases significantly, thus leading to a radiative heat flux near $\omega_3$ that is much smaller for $\varepsilon'_{A2} = 12$ than for $\varepsilon'_{A2} = 1$ (see Fig. S1(b)). As a result, the proportion of the total heat flux due to the high SPhP frequency modes, $\omega_1$ and $\omega_2$, increases as the dielectric function of the substrate increases. This causes an enhancement of the rectification efficiency with $\varepsilon'_{A2}$. A maximum rectification efficiency of 87.2% (rectification factor $R$ of 6.81) occurs at $\varepsilon'_{A2} = 12$ and $T_h$ = 900 K.

[Insert Figure 3]

The impact of the film thickness to gap size ratio $D$ on the rectification efficiency for fixed values $d =10$ nm and $\varepsilon'_{A2} = 12$ is shown in Fig. 4(a). In the temperature band from 700 K to 1000 K, high rectification efficiency is obtained for $D$ values between 0.01 and 0.3. The highest rectification efficiency is 87.7% (rectification factor $R$ of 7.13) when $D = 0.08$ and $T_h = 900$ K. At this temperature, rectification efficiencies of approximately 80% and 70% can be achieved with $D$ values up to 0.2 and 0.3, respectively. SPhP dispersion relation at $T_h = 900$ K, shown in Fig. 4(b), reveals that the two high frequency modes $\omega_1$ and $\omega_2$ lead to strong on- and off-resonance switching with respect to the direction of the temperature bias for $D$ values between 0.01 and 0.3. This is in good agreement with the high rectification efficiency observed in Fig. 4(a). After $D = 0.3$, the rectification efficiency begins to decrease significantly as the on- and off-resonance switching between $\omega_1$ and $\omega_2$ becomes less pronounced. The rectification efficiency vanishes after a $D$ value of 1 because $\omega_1$ and $\omega_2$ are exactly the same in the forward- and reverse-biased scenarios. This behavior can also be observed via the spectral heat flux distributions in Fig. S2 of supplemental material.[27]

[Insert Figure 4]

In summary, this Letter demonstrated that high rectification efficiency can be obtained in a photonic thermal diode with terminals made of the same material supporting SPhPs in the infrared, but with dissimilar structures. It was shown that a specific diode design involving a film and a bulk of 3$C$-SiC can maintain a rectification efficiency in the 80% to 87% range for a wide temperature band ($\sim$ 700 K to 1000 K). Here, the analysis focused on 3$C$-SiC, but the diode concept can utilize any material supporting surface polaritons in the infrared such as doped silicon and silicon dioxide. In addition, while the case of a film and a bulk was treated in this Letter, the rectification efficiency can be potentially further enhanced by exploring more complex nanostructures such as gratings and multilayered media.

**Acknowledgements**

This work was sponsored by the University of Utah Funding Incentive Seed Grant Program.

**Figures**

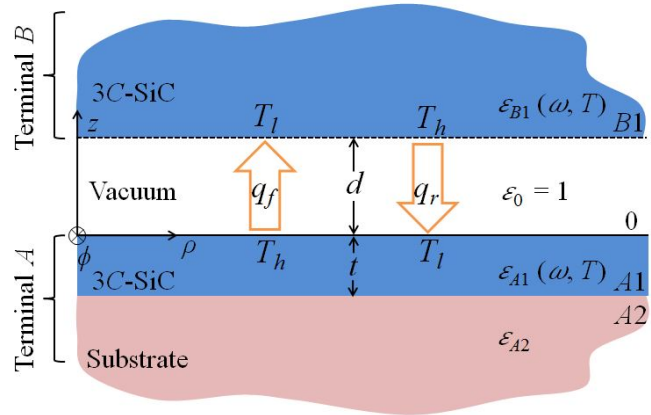

FIG. 1. Schematic of the photonic thermal diode under consideration, where the terminals are separated by a vacuum gap of size $d$. Terminal $A$ is made of a film of $3C$-SiC with thickness $t$ coated on a substrate. Terminal $B$ consists of a bulk of $3C$-SiC.

**(a)**

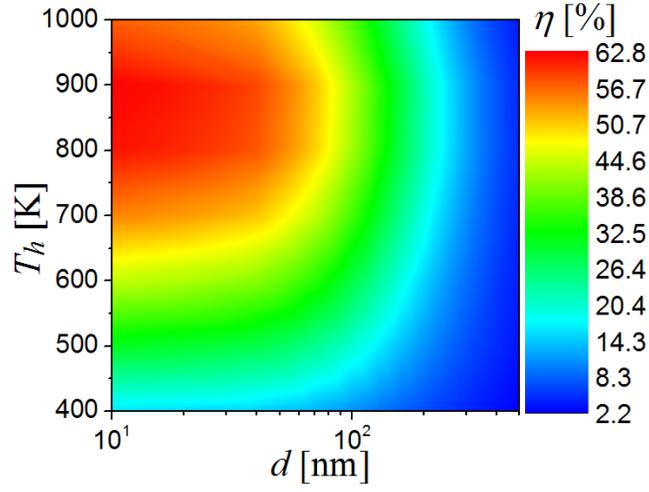

**(b)**

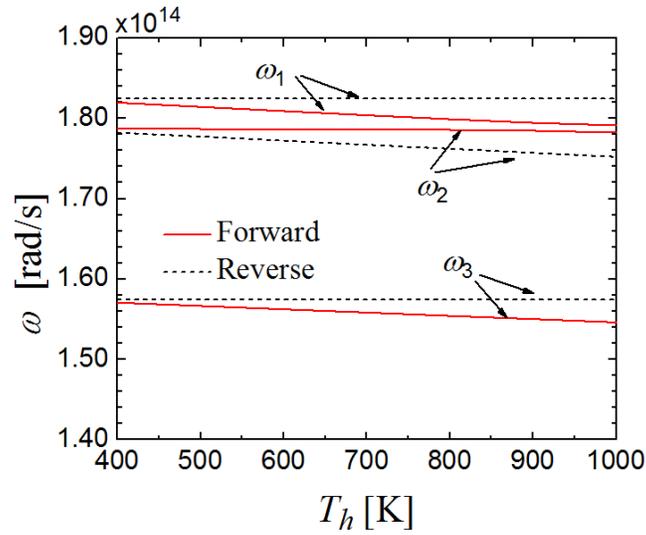

FIG. 2. (a) Rectification efficiency $\eta$ as a function of the temperature $T_h$ and the gap size $d$ for $\varepsilon'_{A2} = 1$ and $D = 0.1$. (b) SPhP dispersion relation for forward- and reverse-biased scenarios as a function of the temperature $T_h$ for $\varepsilon'_{A2} = 1$, $d = 10$ nm and $D = 0.1$.

**(a)**

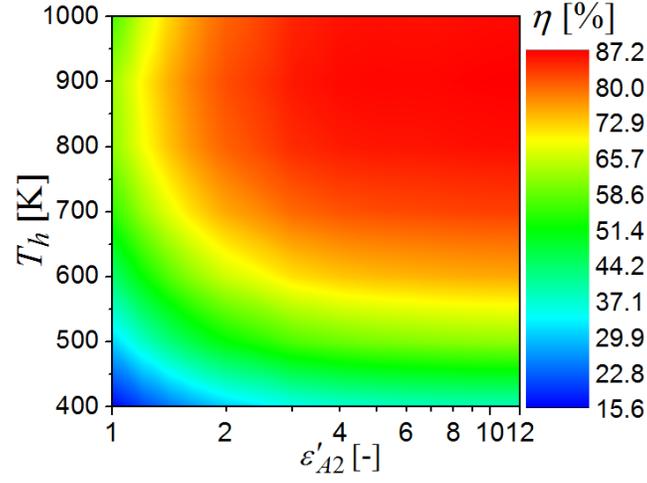

**(b)**

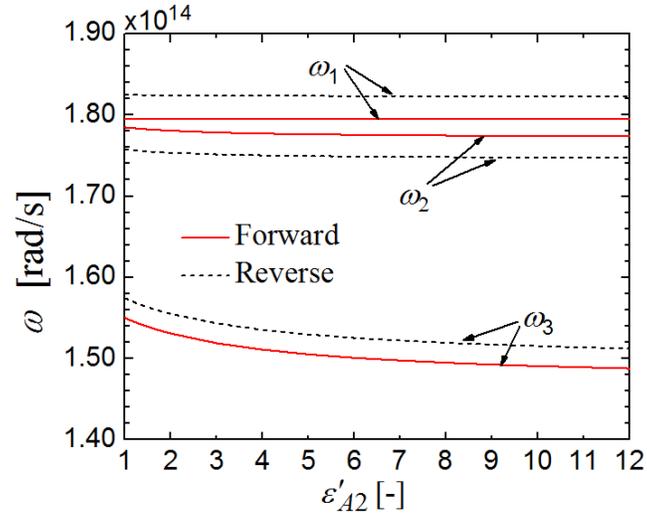

FIG. 3. (a) Rectification efficiency $\eta$ as a function of the temperature $T_h$ and the dielectric function of the substrate $\varepsilon'_{A2}$ for $d$ = 10 nm and $D$ = 0.1. (b) SPhP dispersion relation for forward- and reverse-biased scenarios as a function of dielectric function of the substrate $\varepsilon'_{A2}$ for $T_h$ = 900 K, $d$ = 10 nm and $D$ = 0.1.

**(a)**

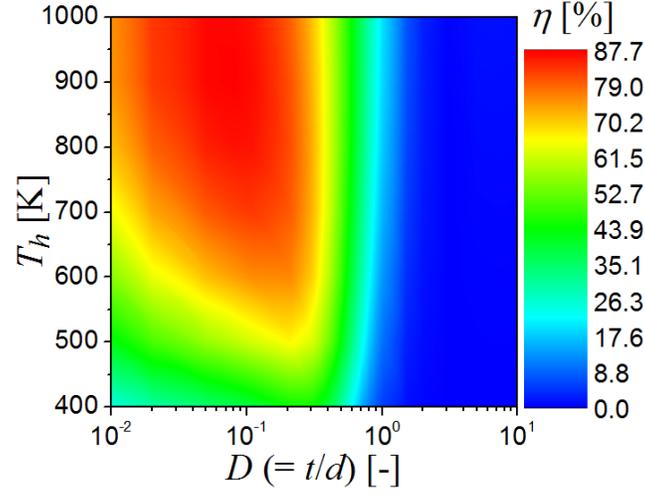

**(b)**

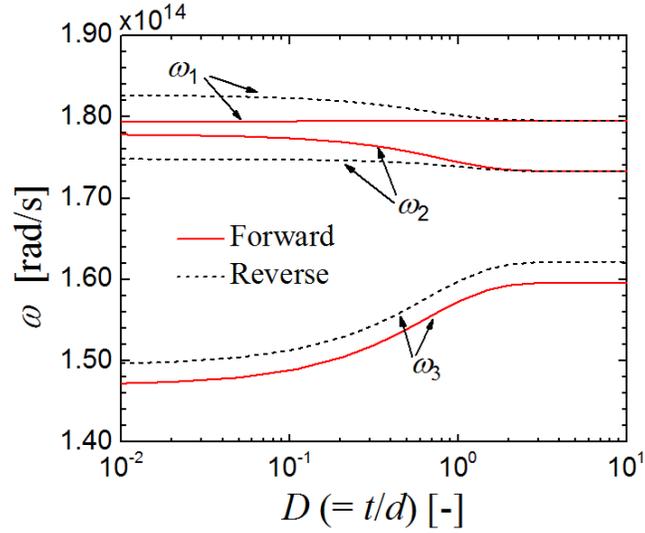

FIG. 4. (a) Rectification efficiency $\eta$ as a function of the temperature $T_h$ and the film thickness to gap size ratio $D$ for $\varepsilon'_{A2} = 12$ and $d = 10$ nm. (b) SPhP dispersion relation for forward- and reverse-biased scenarios as a function of the film thickness to gap size ratio $D$ for $\varepsilon'_{A2} = 12$, $T_h = 900$ K and $d = 10$ nm.

# Supplemental material to Photonic thermal diode enabled by surface polariton coupling in nanostructures

Lei Tang and Mathieu Francoeur[*]

*Radiative Energy Transfer Lab, Department of Mechanical Engineering, University of Utah, Salt Lake City, UT 84112, USA*

## I. SUPPLEMENTAL MATERIAL 1: TEMPERATURE-DEPENDENT DIELECTRIC FUNCTION OF 3*C*-SiC

The temperature-dependent dielectric function of 3*C*-SiC is given by:[21]

$$\varepsilon(\omega, T) = \varepsilon_\infty \left( \frac{\omega^2 - \omega_{LO}^2 + i\Gamma\omega}{\omega^2 - \omega_{TO}^2 + i\Gamma\omega} \right) \tag{S1}$$

where

$$\varepsilon_\infty = 6.7 \exp\left[ 5 \times 10^{-5}(T - 300) \right] \tag{S2a}$$

$$\omega_{LO} = 182.7 \times 10^{12} - 5.463 \times 10^9 (T - 300) \tag{S2b}$$

$$\omega_{TO} = 149.5 \times 10^{12} - 4.106 \times 10^9 (T - 300) \tag{S2c}$$

$$\Gamma = 6.6 \times 10^{11} \left[ 1 + \frac{2}{\exp\left( \dfrac{\hbar\omega_{TO}}{2k_b T} \right) - 1} \right] \tag{S2d}$$

---
[*] Corresponding author. Tel.: +1 801 581 5721, Fax: +1 801 585 9825
Email address: mfrancoeur@mech.utah.edu

In Eq. (S2d), $\hbar$ is the reduced Planck constant and $k_b$ is the Boltzmann constant.

## II. SUPPLEMENTAL MATERIAL 2: SPECTRAL DISTRIBUTIONS OF RADIATIVE HEAT FLUX BETWEEN TERMINALS *A* AND *B* IN THE FORWARD- AND REVERSE-BIASED SCENARIOS

Figure S1 shows spectral distributions of radiative heat flux for forward- and reverse-biased scenarios for $T_h$ = 900 K, $d$ = 10 nm, $D$ = 0.1 and two substrate dielectric functions ($\varepsilon'_{A2}$ = 1 and 12).

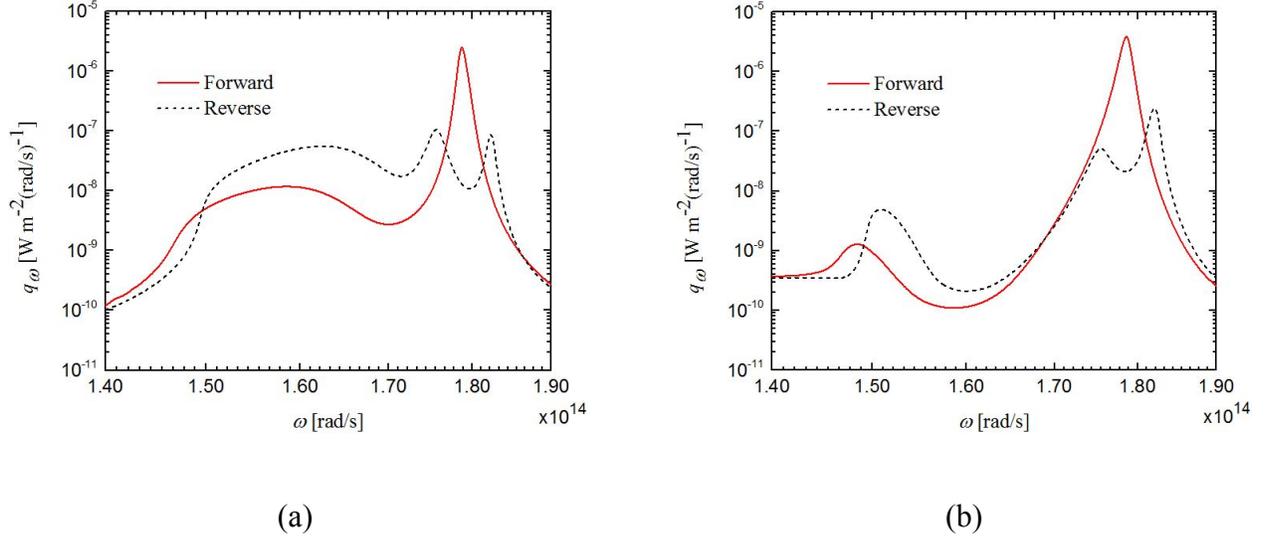

(a)                                                              (b)

FIG. S1. Spectral distributions of radiative heat flux for forward- and reverse-biased scenarios ($T_h$ = 900 K, $d$ = 10 nm, $D$ = 0.1): (a) $\varepsilon'_{A2}$ = 1. (b) $\varepsilon'_{A2}$ = 12.

Figure S2 shows spectral distributions of radiative heat flux for forward- and reverse-biased scenarios for $\varepsilon'_{A2} = 12$, $T_h = 900$ K, $d = 10$ nm and two film thickness to gap ratios ($D = 0.2$ and 1).

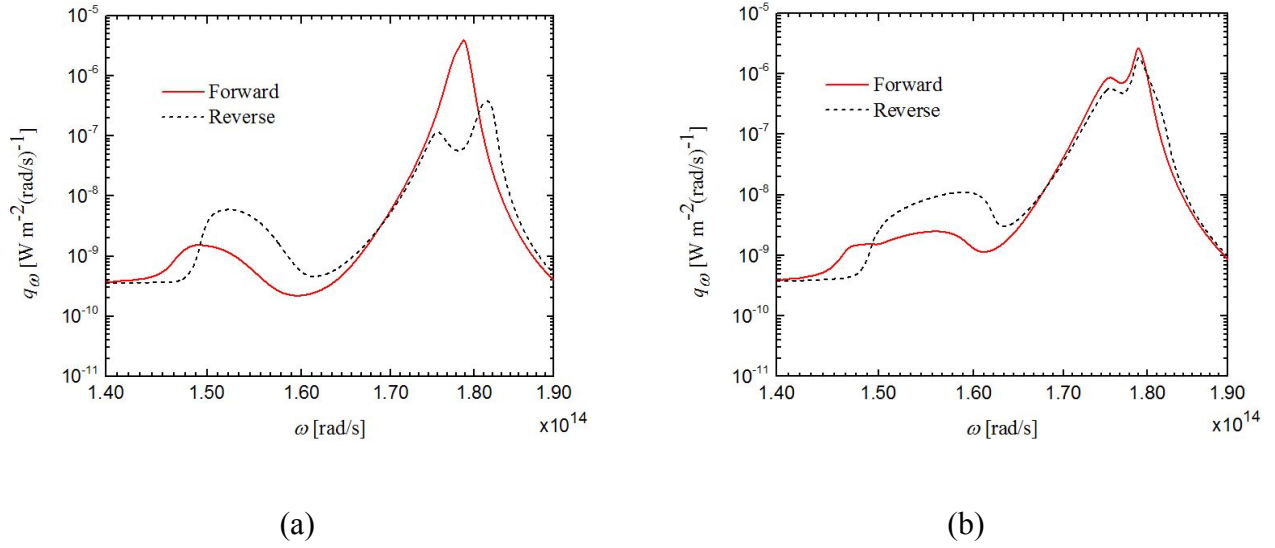

(a)                                                    (b)

FIG. S2. Spectral distributions of radiative heat flux for forward- and reverse-biased scenarios ($\varepsilon'_{A2} = 12$, $T_h = 900$ K, $d = 10$ nm): (a) $D = 0.2$. (b) $D = 1$.